\documentclass[a4paper,10pt,oneside,preprint,5p,number,sort&compress,lefttitle]{elsarticle}

\usepackage{amsfonts,amssymb,amsmath} 
\usepackage{amsthm}         
\usepackage{tensor}
\usepackage{braket}
\usepackage{graphicx} 
\usepackage{xcolor}
\usepackage[colorlinks=true,citecolor=blue,filecolor=black,linkcolor=red,urlcolor=blue]{hyperref}

\usepackage[normalem]{ulem}

\begin{document}

\begin{frontmatter}
\title{Relativistic spin operator must be intrinsic\tnoteref{t1}}
\tnotetext[t1]{This article is registered under preprint number: arXiv:2008.01308}

\author[1]{E. R. F. Taillebois\corref{cor1}}
\ead{emile.taillebois@ifgoiano.edu.br}

\author[2]{A. T. Avelar}
\ead{ardiley@gmail.com}

\cortext[cor1]{Corresponding author}

\address[1]{Instituto Federal Goiano - Campus Avan\c{c}ado Ipameri, 75.780-000, Ipameri, Goi\'{a}s, Brazil}

\address[2]{Instituto de F\'{i}sica, Universidade Federal de Goi\'{a}s, 74.001-970, Goi\^{a}nia, Goi\'{a}s, Brazil}

\begin{abstract}
Although there are many proposals of relativistic spin observables, there is no agreement about the adequate definition of this quantity. This problem arises from the fact that, in the present literature, there is no consensus concerning the set of properties that such an operator should satisfy. Here we present how to overcome this problem by imposing a condition that everyone should agree about the nature of the relativistic spin observable: it must be intrinsic. The intrinsicality concept is analyzed in the relativistic classical limit and then it is extended to the quantum regime, the spin problem being treated in the context of the irreducible unitary representations of the Poincar\'{e} group. This approach rules out three-vector proposals of relativistic spin observable and leads to a unique satisfactory spin definition that, besides being intrinsic, also possesses interesting physical features such as covariance and consistency of predictions in the non relativistic limit. To support the presented results from an operational perspective, a consistent observer-independent model for the electromagnetic-spin interaction is also presented.
\end{abstract}


\end{frontmatter}


\section{Introduction}

To construct mathematical models or theories corresponding to elements of the physical world is one of the main goals of scientists in general and the physicist in particular. In this process, a theoretical construction is considered more fundamental the greater the class of phenomena it is able to describe. Of all models used to describe the physical world, those based on the space-time concept are among the most successful ones, even though some important questions are still open. One of these is the adequate definition of the quantum relativistic spin operator \cite{Frenkel1926,Thomas1927,Bargmann1935,Lubanski1942,Pryce1948, Fleming1965(1), Wouthuysen1950, Czachor1997, Peres2002, Terno2003, Peres2004, Czachor2003, Polyzou2012, Saldanha2012(1), Caban2013, Taillebois2013, Bauke2014,Giacomini2019}.

The spin concept having been discussed for the first time almost a century ago \cite{Frenkel1926,Thomas1927,Bargmann1935}, one might think that everything important concerning this subject has already been said. However, the description of this concept in a fully relativistic scenario is still an open issue with active discussions, several proposals concerning the definition of a relativistic spin operator and its consequences for setups involving the relativistic measurement of spin having been presented over time. In this sense, we highlight, without the pretension of exhausting the matter, the contributions made by Lubański \cite{Lubanski1942} (with its 4-vector description which embodies the idea of internal angular momentum, in spite of difficulties concerning the connection to real measurement setups), Pryce and Fleming \cite{Pryce1948,Fleming1965(1)} (which made explicit the ambiguities that arise when a relativistic position approach is adopted to describe the system’s spin), Foldy and Wouthuysen \cite{Wouthuysen1950} (with their single-particle approach starting from Dirac’s equation), Czachor \cite{Czachor1997} (who presented the first results concerning relativistic quantum information theory), Peres and Terno \cite{Peres2004} (which demonstrated the apparent non-covariance of spin reduced density matrices), Vedral and Saldanha \cite{Saldanha2012(1)} together with Palmer et al. \cite{Palmer2012} (that approached the problem through a relativistic description of a Stern-Gerlach measurement) and ourselves \cite{Taillebois2013} (with an effective approach to spin based on the fiber bundle character of the relativistic momentum-spin Hilbert space). More recently, we emphasize the contribution from Giacomini et al. in \cite{Giacomini2019}, where the recent formalism of relativistic quantum references frames \cite{Giacomini2019n} has been used to address the problem. On the one hand, this problem is relevant to the foundations of physics, since it is directly connected to the problem of relativistic particle localization \cite{Hegerfeldt1974,Zuben2000,Caban2014,Celeri2016}. On the other hand, solving it may be useful to a wide range of research fields, such as relativistic quantum information theory \cite{Gingrich2002,Caban2005,Caban2006,Landulfo2009,Debarba2012,Saldanha2012(2)}, quantum spintronics \cite{Awschalom2013}, and light-matter interaction at relativistic intensities \cite{Piazza2012, Ahrens2012}.

Here it is argued that the origin of the relativistic spin operator controversy is the lack of a satisfactory definition of what is meant by intrinsicality. Thus, despite the question we have to eventually answer be ``Which spin operator is actually connected to a realistic relativistic spin measurement?", a theoretical approach of what is an intrinsic property in special relativity is exposed first. Once this is done, a unique intrinsic relativistic spin operator and its connection to a realistic experimental setup are presented. This approach allows to shed light over the properties that a spin observable has to satisfy and some misconceptions concerning intrinsicality.

The article is organized as follows.  In Section \ref{sec:Int} we introduce the concept of intrinsicality, its mathematical implementation being formalized for the particular case of quantities defined in the context of special relativity. Assuming the intrinsic character as fundamental for an adequate definition of the relativistic spin concept, in Section \ref{sec:Spin} it is shown that three-vector descriptions of such a quantity must be ruled out. Besides that, in this same section, intrinsicality is also used to define a unique satisfactory intrinsic relativistic spin observable, a consistent observer-independent model for the electromagnetic-spin interaction being presented in Section \ref{sec:Elec}. Finally, final remarks and conclusions are presented in Section \ref{sec:Conc}.

\section{\label{sec:Int}Intrinsicality}

In general the definition of a physical property, such as position or momentum, may be split into three stages. First, the concept of interest $\mathcal{A}$ must be defined, introducing what the property should represent. After this, it is necessary to associate an appropriate mathematical structure $M(\mathcal{A})$ to the concept. The last step is the description: it is necessary to quantify the mathematical structure. This is usually done by introducing a basis $\mathcal{O}$ associated to an observer that measures the property, the description being denoted here by $D(M(\mathcal{A}), \mathcal{O})$. Thus, a specific concept $\mathcal{A}$ leads to a mathematical structure $M(\mathcal{A})$, that is quantified in a basis $\mathcal{O}$ by a description $D(M(\mathcal{A}), \mathcal{O})$.

It is worth nothing that, although the description in general depends on the observer, the mathematical structure may also be observer dependent, as pointed by Fleming \cite{Fleming1965(1)} in the context of quantum relativistic position operators. A simple example of a property with such a behavior is the orbital angular momentum of a set of particles in classical mechanics: if the origin of the system of coordinates is changed, the mathematical structure that describes this property also changes, implying that this property is both system and observer dependent. The independence of the mathematical structure relative to the observer is at the heart of the intrinsic character of a property.

The mathematical definition of a physical property in Minkowski space-time is presented in what follows, Penrose's abstract index tensor notation \cite{Malcolm} being used whenever the introduction of a basis is unnecessary. 

\subsection{Physical property in Minkowski space-time}

\begin{figure}[ht]
\centering
\includegraphics[scale=0.8]{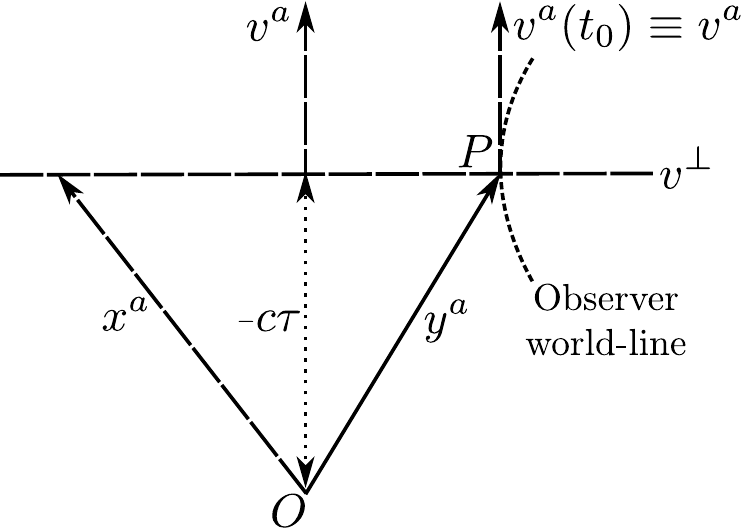}
\caption{\footnotesize{Measurement hyperplane $v^{\perp}$ related to an observer of instantaneous 4-velocity $v^{a}(t_{0}) = v^{a}$. The hyperplane is composed by all events that satisfies the relation $x^{a}v_{a} = -c^{2}\tau$, $x^{a}$ describing their position 4-vectors and $\tau$ an invariant parameter that gives the position of the hyperplane relative to the origin $O$. The point $P$ of intersection between the observer world line and the measurement hyperplane is described by the position 4-vector $y^{a}$.}}
\label{Fig1}
\end{figure}

Suppose that an observer of 4-velocity $v^{a}(t)$, where $t$ is his proper time, wants to analyze some property $\mathcal{A}$ of a classical system at a given instant $t_{0}$. A non-trivial lecture of this statement is that the measurement is being done relative to a space-like hyperplane $v^{\perp}$ that intercepts the observer world-line at a point $P$ where his 4-velocity is $v^{a}(t_{0}) \equiv v^{a}$. If an arbitrary space-time point $O$ is chosen as reference, the later statement implies that the property is being defined relative to an hyperplane given by
\begin{equation}
x^{a}v_{a} = -c^{2}\tau,
\label{hiper}
\end{equation}
$\tau$ being an invariant parameter that defines the position of $v^{\perp}$ relative to $O$ (FIG. \ref{Fig1}). Since the mathematical structure $M(\mathcal{A})$ associated by the observer to the property $\mathcal{A}$ may depend on the space-like hyperplane of measurement and the location of $P$ in that hyperplane, from now on it will be denoted by $M(\mathcal{A}) \equiv A(v^{a},\tau,y^{b})$, $y^{b}$ being the position four-vector of $P$ relative to $O$.

\begin{figure}[ht]
\centering
\includegraphics[scale=0.85]{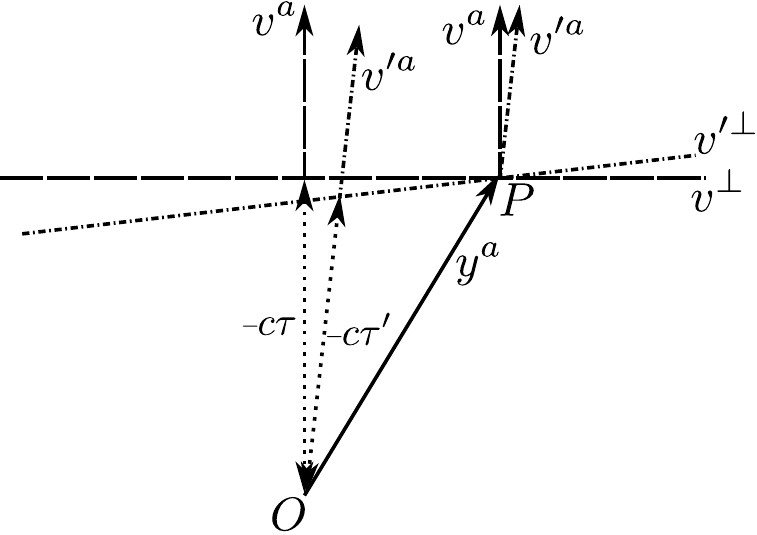}
\caption{\footnotesize{Measurement hyperplanes associated to different observers that pass by the event $P$ described by the position 4-vector $y^{a}$ relative to the event of reference $O$. The first (second) observer of instantaneous 4-velocity $v^{a}$ ($v^{\prime a}$) defines a property relative to the hyperplane $v^{\perp}$ ($v^{\prime \perp}$) that is fixed by the observer 4-velocity and the invariant parameter $\tau$ ($\tau^{\prime}$).}}
\label{Fig2}
\end{figure}

If a second observer, whose world-line also passes by $P$, decides to measure this same property, the mathematical structure that he will associate to the property is going to be related, in general, to an hyperplane that depends on his 4-velocity $v^{\prime a}$ at $P$ (FIG. \ref{Fig2}). Thus, the property as seen by the second observer is related to the hyperplane $v^{\prime \perp}$ given by
$x^{a}v^{\prime}_{a} = -c^{2}\tau^{\prime},$
and the associated mathematical structure is of the form $A(v^{\prime a}, \tau^{\prime}, y^{b})$.

Once established this behavior for a general mathematical structure associated to an arbitrary property, the concept of intrinsicality can be introduced. A property of a system is intrinsic if the related mathematical structure does not depend on the observer that is measuring it (although the description of the property may depend on the choice of a specific observer dependent basis). Thus, the definition of an intrinsic property in special relativity is associated to the dependence of the mathematical structure on the hyperplane related to which it is defined: the mathematical structure of an intrinsic property cannot depend on the orientation of the hyperplane on which it is measured (punctual Lorentz invariance) or on the point of intersection $P$ of the observer world-line with the measurement hyperplane (space-like translational invariance).

\subsection{Space-like translational invariance}

An arbitrary infinitesimal variation of the observer's world line produces a change in $A(v^{a},\tau,y^{b})$ given by
\begin{equation*}
\delta A(v^{a},\tau,y^{b}) = \left(\delta v^{c}\frac{\partial}{\partial v^{c}} + \delta \tau\frac{\partial }{\partial \tau} + \delta y^{c}\frac{\partial}{\partial y^{c}}\right)A(v^{a},\tau,y^{b}).
\end{equation*}
To establish the consequences of space-like translational invariance, an infinitesimal translation of $P$ inside $v^{\perp}$ is considered. This corresponds to setting $\delta v^{c} = 0$ and $\delta \tau = 0$, which leads the constraint \eqref{hiper} to $v_{c}\delta y^{c} = 0$, resulting in
\begin{equation}
\delta A(v^{a},\tau,y^{b}) = \left(\frac{\partial A(v^{a},\tau,y^{b})}{\partial y^{c}} + \lambda v_{c}\right)\delta y^{c} = 0,
\label{link1}
\end{equation} 
where $\lambda$ is a Lagrange multiplier introduced to take into account the constraint. Solving \eqref{link1} for $\lambda$ implies that
\begin{equation}
\left(\frac{\partial}{\partial y^{c}} + v_{c}\frac{v^{d}}{c^{2}}\frac{\partial}{\partial y^{d}}\right) A(v^{a},\tau,y^{b}) = 0
\label{vinc1}
\end{equation}
for a property to be invariant under observer translations inside $v^{\perp}$.

\subsection{Punctual Lorentz invariance}

For the case of punctual Lorentz invariance relative to $P$, $\delta y^{c} = 0$ and the hyperplane constraint \eqref{hiper} implies that $y_{c}\delta v^{c} = -c^{2} \delta \tau$. In addition, this kind of transformation does not change $v^{c}$ in an arbitrary way, since the relation $v_{c}v^{c} = -c^{2}$ cannot be violated. Then, the invariance condition must be written as
\begin{eqnarray}
\delta A(v^{a},\tau,y^{b}) & = & \left(\frac{\partial A(v^{a},\tau,y^{b})}{\partial v^{c}} + \lambda_{1}y_{c} + \lambda_{2}v_{c}\right)\delta v^{c} \nonumber \\
& + & \left(\frac{\partial A(v^{a},\tau,y^{b})}{\partial \tau} + \lambda_{1}c^{2}\right)\delta \tau = 0,
\label{link2}
\end{eqnarray}
where $\lambda_{1}$ and $\lambda_{2}$ are Lagrange multipliers introduced to take into account the hyperplane and 4-velocity norm invariance. Solving \eqref{link2} for $\lambda_{1}$ and $\lambda_{2}$ leads to
\begin{equation}
\left(\frac{\partial}{\partial v^{c}} + \frac{v_{c}v^{d}}{c^{2}}\frac{\partial}{\partial v^{d}} - \frac{(y_{c} - \tau v_{c})}{c^{2}}\frac{\partial}{\partial \tau}\right) A(v^{a},\tau,y^{b}) = 0
\label{vinc2}
\end{equation}
for a property to be independent of the measurement hyperplane orientation. 

\subsection{Intrinsic physical property in Minkowski space-time}

Relations \eqref{vinc1} and \eqref{vinc2} describe, respectively, the space-like translational invariance and the punctual Lorentz invariance. However their mutual implications also must be taken into account. Performing a punctual Lorentz variation in \eqref{vinc1} and taking \eqref{vinc2} into account results in $A(v^{a},\tau, y^{b}) \equiv A(v^{a}, y^{b})$, which allows to rewrite \eqref{vinc2} as
\begin{equation}
A(v^{a} + \delta v^{a},y^{b})\Big|_{v_{a}\delta v^{a} = 0} = A(v^{a},y^{b})
\label{res1}
\end{equation}
On the other hand, performing a space-like translation variation in \eqref{res1} and taking \eqref{vinc1} into account, it follows that
\begin{equation}
A(v^{a},y^{b}) \equiv A(v^{a})
\label{trans}
\end{equation}
and Equation \eqref{res1} may be rewritten as
\begin{equation}
A(v^{a} + \delta v^{a})\Big|_{v_{a}\delta v^{a} = 0} = A(v^{a}).
\label{res4}
\end{equation}
Equation \eqref{trans} states that to be intrinsic the property must be invariant under any arbitrary translation of the point of observation, while equation \eqref{res4} states the invariance under arbitrary Lorentz transformations related to an arbitrary point of reference. Thus, a relativistic intrinsic property must be associated to the same mathematical structure independent of the observer that is performing the measurement.

\section{\label{sec:Spin} Intrinsicality applied the problem of a relativistic spin operator}

Starting from the framework of intrinsicality, the problem of finding a satisfactory intrinsic spin operator can now be addressed. However, it is instructive to analyze first if there are proposals of relativistic spin operators present in literature that can be ruled out by assuming that the intrinsic character must be fundamental for an observable to be called spin. 

\subsection{Non-intrinsic relativistic spin operator}

\begin{figure}[ht]
\centering
\includegraphics[scale=0.9]{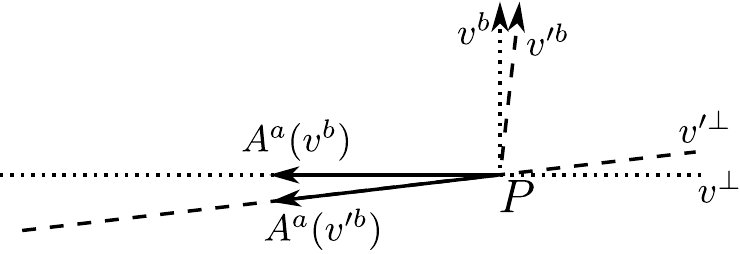}
\caption{\footnotesize{Dependence of a 3-vector property on the measurement hyperplane. If $\mathcal{A}$ is a property that is described as a 3-vector by any observer, then the associated mathematical structure will be observer dependent. For an observer of 4-velocity $v^{b}$ it will be given by $A^{a}(v^{b}) \in v^{\perp}$ while for an observer of 4-velocity $v^{\prime b}$ it will be given by $A^{a}(v^{\prime b}) \in v^{\prime \perp}$.}}
\label{Fig3}
\end{figure}

It is usual to require that to be a spin observable an operator must be a 3-vector, and some of the most important relativistic spin proposals present in literature satisfy this condition \cite{Terno2003,Peres2002,Peres2004,Czachor1997,Giacomini2019}. However, in special relativity a 3-vector can be defined only if a particular space-like hyperplane is specified. Thus, if all observers that pass through a space-time point $P$ define a property as being a 3-vector, the mathematical structure will be observer dependent, since each observer may be associated to a different space-like measurement hyperplane (FIG. \ref{Fig3}). This indicates that a property that is always represented as a 3-vector quantity by any observer cannot be an intrinsic property. A formal deduction of this statement is presented in what follows.

A 3-vector property defined by an observer of 4-velocity $v^{a}$ can be represented by a 4-vector structure $A^{a}(v^{b})$ that satisfies
\begin{equation}
A^{a}(v^{b})v_{a} = 0.
\label{oi1}
\end{equation}
Since it is assumed that any observer passing by the measurement point associates a 3-vector to the property, relation \eqref{oi1} must hold for any four-velocity, i.e.
\begin{equation}
A^{a}(v^{b} + \delta v^{b})(v_{a} + \delta v_{a})\Big|_{v^{b}\delta v_{b} = 0} = 0.
\label{oi2}
\end{equation}
For such a property to be intrinsic it must satisfy \eqref{res4} and, using this together with \eqref{oi1} and \eqref{oi2}, lead to
\begin{equation}
A^{a}(v^{b})\delta v_{a}\Big|_{v^{b}\delta v_{b} = 0} = 0.
\label{oi3}
\end{equation}
Hence, equations \eqref{oi1} and \eqref{oi3} impose that $A^{a}(v^{b}) = 0$, implying that only the null 3-vector can be associate to an intrinsic relativistic property. This result has strong implications, since it invalidates the use of three-vector operators as spin observables if this property is expected to be intrinsic. Besides that, it makes explicit that the problem of non covariance of spin observables present in the relativistic quantum information theory literature \cite{Czachor1997,Peres2002,Terno2003,Peres2004,Landulfo2009} is in fact a consequence of the  non intrinsicality of the observables considered there.

\subsection{Intrinsic relativistic spin operator}

Since three-vectors cannot define an intrinsic property in relativity theory, an important question is left open: is there in fact an intrinsic relativistic spin operator? To answer it, the framework of irreducible unitary representations of the Poincar\'{e} group \cite{Wigner} will be adopted. The advantage of using this formalism is that the results are not restricted only to spin-half particles, being valid for any spin value compatible with the Poincar\'{e} symmetry. A thorough explanation of how this formalism is connected to Dirac spinor formalism for spin-half particles can be found in \cite{Caban2005}.

Quantum mechanically the mathematical structure associated to a property is an observable operator and, on the basis of the correspondence principle \cite{Fleming1965(1)}, the classical results present in \eqref{trans} and \eqref{res4} can be mapped to quantum mechanics. As a result, a property that is intrinsic in relativistic quantum mechanics must be associated to a mathematical structure given by an operator $A(v^ {a})$ that satisfies
\begin{equation}
\mathrm{Tr}\left(\rho A(v^{a})\right) = \mathrm{Tr}\left(\rho^{\prime} A(v^{a} + \delta v^{a})\Big|_{v_{a}\delta v^{a} = 0}\right),
\label{intrinsic1}
\end{equation}
$\rho$ and $\rho^{\prime}$ being the density matrices of the system as described by the two different observers. In the same way, the result concerning the impossibility of using 3-vectors to describe relativistic intrinsic properties also applies to observables in relativistic quantum mechanics.

To determine if there is an intrinsic spin observable, it is necessary to define primarily what is the concept that this property must describe. This is done by imposing a set of three physical reasonable assumptions, the first of them being intrinsicality. Since the spin concept is widely understood as the intrinsic angular momentum of the system, it is natural to define this quantity as being the intrinsic part of the second rank antisymmetric tensor $J_{ab}$ that describes the relativistic angular momentum concept. This construction allows to split the total angular momentum $J_{ab}$ into an orbital $L_{ab}$ and a spin $S_{ab}$ part, both being second rank antisymmetric objects: $J_{ab} = L_{ab} + S_{ab}$. Since $S_{ab}$ must be intrinsic, Equation \eqref{intrinsic1} leads to
\begin{equation}
\begin{split}
\left[ J_{\alpha\beta}, S_{\mu\nu} \right] & = 4i\eta_{[\alpha[\nu}S_{\mu]\beta]}, \\
\left[ P_{\alpha}, S_{\mu\nu} \right] & = 0.
\end{split} \label{eq:dfg}
\end{equation}
These results mean that intrinsicality together with the second rank character imply that $S_{ab}$ must transform as a second rank tensor under arbitrary Poincar\'{e} transformations.

Secondly, the 1-form operator defined by
\begin{equation*}
\tensor[^*]{S}{_{ab}}P^{b} = -\frac{1}{2}\varepsilon_{abcd}S^{cd}P^{b}
\end{equation*}
must coincide with $W_{a} = \tensor[^{*}]{J}{_{ab}}P^{b}$, the so-called Pauli-Lubanski operator. To understand that assumption lets analyze the angular momentum operator $J_{ab}(p)$ for a particle of definite momentum $p^{a}$. For an inertial observer $\mathcal{O}_{v}$ of 4-velocity $v^{a}$, the information contained in $J_{ab}(p)$ can also be retrieved using the pair of 1-forms
\begin{subequations}
\begin{eqnarray}
J_{a}(p;v) & = & \tensor[^*]{J}{_{ab}}\frac{v^{b}}{c} \label{eq2} \\
K_{a}(p;v) & = & J_{ab}\frac{v^{b}}{c}, \label{eq1}
\end{eqnarray}
\end{subequations}
their connection with $J_{ab}(p)$ being given by
\begin{equation}
J_{ab}(p) = \frac{2}{c}v_{[a}K_{b]}(p;v) + \frac{1}{c}\varepsilon_{abcd}v^{c}J^{d}(p;v).
\label{oi10}
\end{equation}
The operators $J_{a}(p;v)$ and $K_{a}(p;v)/c$ are interpreted, respectively, as the total angular and mass momentum observables for a particle of 4-momentum $p^{a}$ as seen by the observer $\mathcal{O}_{v}$, and
\begin{equation}
K_{a}(p;v)v^{a} = J_{a}(p;v)v^{a} = 0. \label{eq10}
\end{equation}
Assuming now that the inertial observer has the same 4-velocity of the particle, i.e. $v^{a} = p^{a}/m$, where $m$ is the particle's mass, equations \eqref{eq2} and \eqref{eq1} can be written as
\begin{subequations}
\begin{eqnarray}
J_{a}(p;p/m) & \equiv & J_{a}(p) = \tensor[^*]{J}{_{ab}}\frac{p^{b}}{mc} \label{eq4} \\
K_{a}(p;p/m) & \equiv & K_{a}(p) = J_{ab}\frac{p^{b}}{mc}. \label{eq3}
\end{eqnarray}
\end{subequations}
Comparing \eqref{eq4} with the Pauli-Lubanski operator and using the interpretation presented before, $J_{a}(p) = \frac{W_{a}(p)}{mc}$ is the total angular momentum of the particle as seen by a co-moving observer. From non-relativistic quantum mechanics it is known that the property called \textit{spin} must be related to the total angular momentum of the particle in its rest frame, thus, assuming that $S_{ab}(p)$ is the extension of this concept, it is expected that the angular part of $S_{ab}(p)$ taken in the rest frame of the particle possesses this interpretation, such that $W_{a}(p) = \tensor[^{*}]{S}{_{ab}}(p)p^{b}$. Extending this to an arbitrary momentum state leads to the second assumption that
\begin{equation}
W_{a} = \tensor[^{*}]{S}{_{ab}}P^{b}.
\label{cond10}
\end{equation}

However, since $W_{a}/mc$ does not contain the full information presented in $S_{ab}$, it is necessary to define the extra component $S_{ab}P^{b}/mc$ to fully determine $S_{ab}$. To do so, a last assumption is made: the intrinsic spin observable must be consistent in the non-relativistic limit, implying that it must lead to a single 3-vector angular momentum quantity in that limit. This means that in the rest frame of the particle it is expected that only the angular momentum contribution of $S_{ab}$ must persist. This, together with \eqref{eq10}, imposes that
\begin{equation}
S_{ab}P^{b}/mc = 0.
\label{cond2}
\end{equation}
Due to the tensor character of $S_{ab}$, which derived from the imposition of intrinsicality, the same construction presented in \eqref{oi10} can be used for $S_{ab}$ and, using \eqref{cond10} and \eqref{cond2}, leads to
\begin{equation}
S_{ab} = \frac{i}{P^{2}}[W_{a},W_{b}].
\label{intspin}
\end{equation}
Equation \eqref{intspin} corresponds to the unique observable that is intrinsic and satisfies the set of minimal conditions presented to define what it is expected from a spin observable. Besides that, $S_{ab}S^{ab}/2 = s(s+1)$, where $s$ is the spin of the particle, i.e. the usual spin norm is frame invariant.

To fully understand the implications of assuming that \eqref{intspin} is the only satisfactory description of an intrinsic spin observable, lets decompose this operator in the 1-form observer dependent description. For an inertial observer of 4-velocity $v^{a}$, the information contained in \eqref{intspin} can be described by the couple of observables $\Sigma_{a}(v) = \tensor[^*]{S}{_{ab}}v^{b}/c$ and $M_{a}(v) = S_{ab}v^{b}/c$. The observables $\Sigma_{a}(v)$ and $M_{a}(v)$ correspond, respectively, to the angular and mass momentum part of $S_{ab}$ as seen by an observer of 4-velocity $v^{a}$. It is important to stress that separately these quantities are not intrinsic properties, since the information that each one carries is observer dependent.

Using a basis ($e^{a}_{(\alpha)}$) with $e^{a}_{(0)} = v^{a}/c$, observable $\Sigma_{a}(v)$ can be written as a 3-vector quantity $\mathbf{\Sigma}(v)$ given by
\begin{equation*}
\mathbf{\Sigma}(v) = \frac{1}{mc}\left(P^{0}\mathbf{S} - \frac{(\mathbf{S}\cdot \mathbf{P})\mathbf{P}}{mc + P^{0}}\right) 
\end{equation*} 
where $\mathbf{S}$ is the usual Wigner spin operator \cite{Peres2002,Peres2004}. Although the observable $\mathbf{\Sigma}(v)$ transforms as a 3-vector under the action of rotations inside $v^{\perp}$, it does not satisfy the spin algebra unless $v^{a}$ is a co-moving observer. This fact could be used to argument that $S_{ab}$ is not an adequate spin observable, however this argument is misleading and can be circumvented. The requirement of the spin algebra comes from the fact that it is expected that the spin observable be a generator of the rotation group, thus satisfying the $\mathfrak{su}(2)$ algebra. However, if spin is defined as fundamentally being the intrinsic part of $J_{ab}$, the angular momentum character being imposed only in the rest frame of the particle, there is no reason to state that the $\mathfrak{su}(2)$ algebra must be satisfied in an arbitrary relativistic frame of reference other than the particle rest frame. In fact this behavior is even expected since, except if a co-moving observer is being considered, there is not a passive symmetry transformation that rotates only the particle's spin relative to the observer. This schism between the generator and observable character is not new in physics, being present for example in the splitting of electromagnetic angular momentum operator into spin and orbital parts \cite{Enk1994}.

\section{\label{sec:Elec} Electromagnetic coupling with intrinsic spin}

As pointed at the beginning, a theoretical definition of a relativistic spin observable is incomplete if it is not connected to a realistic relativistic spin measurement. Thus, to complete the theoretical argument given above, a simple model of coupling between $S_{ab}$ and the electromagnetic field tensor $F^{ab}$ is presented. The simplest description for this coupling is given by the interaction Hamiltonian
\begin{equation}
\mathcal{H}_{I} = -\alpha \frac{F^{ab}S_{ab}}{2},
\label{hi}
\end{equation}
where $F^{ab}$ is the classical electromagnetic tensor and $\alpha$ is the gyromagnetic ratio. This is in agreement with what was proposed in \cite{Saldanha2012(1),Saldanha2012(2)}, since \eqref{hi} can be rewritten as $\mathcal{H}_{I} = -\alpha B^{a}(P)W_{a}/mc$, where $B_{a}(P) = \tensor[^*]{F}{_{ab}}P^{b}/mc$ is the operator that gives the magnetic field in the rest-frame of the particle, i.e. the interaction is given by the usual spin-magnetic coupling in the instantaneous rest frame of the particle. Equation \eqref{hi} can also be written in observer's frame as
\begin{equation}
\mathcal{H}_{I} = -\alpha \left[\frac{(\mathbf{E}\times\mathbf{P})\cdot\mathbf{S}}{mc^{2}} + \left(\frac{P^{0}}{mc}\mathbf{S}_{\perp} + \mathbf{S}_{\parallel}\right)\cdot\mathbf{B} \right],
\label{spinfield}
\end{equation}
$\mathbf{E}$ and $\mathbf{B}$ being the electric and magnetic fields in this frame, while $\mathbf{S}_{\parallel} = \mathbf{S}\cdot\mathbf{P}/\|\mathbf{P}\|$ and $\mathbf{S}_{\perp} = \mathbf{S} - \mathbf{S}_{\parallel}$ are the parallel and perpendicular components of $\mathbf{S}$ relative to $\mathbf{P}$. Considering a particle of definite 4-momentum $p^{a}$ in non-relativistic limit, equation \eqref{spinfield} can be rewritten as
\begin{equation}
\mathcal{H}_{I} = -\alpha\left[\frac{(\mathbf{E}\times\mathbf{p})\cdot\mathbf{S}}{mc^{2}} + \mathbf{S}\cdot\mathbf{B}\right].  \label{eq:gui}
\end{equation}
The second term in \eqref{eq:gui} is just the usual magnetic-spin interaction, while the first term is just twice the usual spin interaction of the moving electron with the electric field. The fact that \eqref{hi} has led to twice the correct electric interaction term is expected since the Thomas correction \cite{Thomas1926}, that gives a careful treatment for the contribution of electron's acceleration to the electric interaction, was not taken into account. If such contribution is introduced in the interaction Hamiltonian, the above prescription gives exactly the expected electromagnetic spin interaction in the non-relativistic limit, supporting the use of \eqref{intspin} as the adequate intrinsic spin operator.

\section{\label{sec:Conc} Conclusions}

To summarize, it was shown that the use of three-vectors for the description of relativistic spin is not consistent whit the usual conception that this quantity must be associated to the intrinsic angular momentum of the system, the problem of the so-called non covariance of spin observables \cite{Czachor1997,Peres2002,Terno2003,Peres2004,Landulfo2009} being in fact a consequence of the  non intrinsicality of three-vector observables. Then, starting from the intrinsicality condition, it was shown that there is a unique relativistic spin operator that satisfies the properties expected from the concept of an intrinsic angular momentum quantity.

As a final result, it should be noted that the spin observable \eqref{intspin} obtained by imposing intrinsicality as a fundamental property leads to the point-like center-of-inertia position operator \cite{Pryce1948,Fleming1965(1)}. Indeed, this is remarkable since the center-of-inertia of a single particle is associated to its mass, a property that also satisfies the constraints imposed by intrinsicality. This raise an interesting question concerning the definition of intrinsic properties and the localization problem: are the intrinsic properties of a system always related to the same position operator? The answer to this question is left open, a thorough analysis of it being postponed to a future work.

The authors are thankful for the support provided by the Brazilian agencies CAPES (PROCAD2013), FAPEG (PRONEX \#201710267000503, PRONEM \#2017102670-00540), CNPq (\#459339/2014-1, \#312723/2018-0) and the Instituto Nacional de Ciência e Tecnologia - Informação Quântica (INCT-IQ). We also thank Flaminia Giacomini for bringing references \cite{Giacomini2019,Giacomini2019n} to our attention.



\begin{thebibliography}{10}

\bibitem{Frenkel1926}
J. Frenkel
\newblock {\em Z. Phys.} 37, 243 (1926).

\bibitem{Thomas1927}
L. H. Thomas.
\newblock {\em Phil. Mag.} 3, 1 (1927).

\bibitem{Bargmann1935}
V. Bargmann, L. Michel, and V. L. Telegdi
\newblock {\em Phys. Rev. Lett.} 2, 435 (1935).

\bibitem{Lubanski1942}
J. K. Luba\'{n}ski
\newblock {\em Physica (Utrecht)} 9, 310 (1942).

\bibitem{Pryce1948}
M.~H.~L. Pryce.
\newblock {\em Proc. R. Soc. Lond. A} 195, 62 (1948).

\bibitem{Fleming1965(1)}
G.~N. Fleming.
\newblock {\em Phys. Rev.} 137, B188 (1965).

\bibitem{Wouthuysen1950}
L. L. Foldy and S. A. Wouthuysen
\newblock {\em Phys. Rev.} 78, 29 (1950).

\bibitem{Czachor1997}
M. Czachor.
\newblock {\em Phys. Rev. A} 55, 72 (1997).

\bibitem{Peres2002}
A. Peres, P.~F. Scudo, and D.~R. Terno.
\newblock {\em Phys. Rev. Lett.} 88, 230402 (2002).

\bibitem{Terno2003}
D.~R. Terno.
\newblock {\em Phys. Rev. A} 67, 014102 (2003).

\bibitem{Peres2004}
A. Peres and D.~R. Terno.
\newblock {\em Rev. Mod. Phys.} 76, 93 (2004).

\bibitem{Czachor2003}
M. Czachor and M. Wilczewski.
\newblock {\em Phys. Rev. A} 68, 010302(R) (2003).

\bibitem{Polyzou2012}
W. N.~Polyzou, W.~Glockle, and H.~Witala.
\newblock {\em Few-Body Syst.} 54, 1667 (2012).

\bibitem{Saldanha2012(1)}
P. L. Saldanha and V. Vedral.
\newblock {\em New J. Phys.} 14, 023041 (2012).

\bibitem{Caban2013}
P.~Caban, J.~Rembieli{\'n}ski, and M.~Wlodarczyk.
\newblock {\em Ann. Phys.} 330, 263 (2013).

\bibitem{Taillebois2013}
E.~R.~F. Taillebois and A.~T. Avelar.
\newblock {\em Phys. Rev. A} 88, 060302(R) (2013).

\bibitem{Bauke2014}
{H. Bauke, S. Ahrens, C. H.  Keitel, and R. Grobe}.
\newblock {\em New J. Phys.} 16, 043012 (2014).

\bibitem{Giacomini2019}
{F. Giacomini, E. Castro-Ruiz, and \v{C}. Brukner}.
\newblock {\em Phys. Rev. Lett.} 123, 090404 (2019).

\bibitem{Palmer2012}
M. C. Palmer, M. Takahashi, and H. F. Westman.
\newblock {\em Ann. Phys.} 336, 505 (2012).

\bibitem{Giacomini2019n}
{F. Giacomini, E. Castro-Ruiz, and \v{C}. Brukner}.
\newblock {\em Nat. Comm.} 10, 494 (2019).

\bibitem{Hegerfeldt1974}
G. C. Hegerfeldt.
\newblock {\em Phys. Rev. D} 10, 3320 (1974).

\bibitem{Zuben2000}
F. S. G. Von Zuben.
\newblock {\em J. Math. Phys.} 41, 6093 (2000).

\bibitem{Caban2014}
P.~Caban, J. Rembieli{\'n}ski, P. Rybka, K. A. Smoli{\'n}ski, and P.~Witas.
\newblock {\em Phys. Rev. A} 89, 032107 (2014).

\bibitem{Celeri2016}
L. C. C\'{e}leri, V. Kiosses and D. R. Terno.
\newblock {\em Phys. Rev. A} 94, 062115 (2016).

\bibitem{Gingrich2002}
R. M. Gingrich and C. Adami.
\newblock {\em Phys. Rev. Lett.} 89, 270402 (2002).

\bibitem{Caban2005}
P. Caban and J. Rembieli{\'n}ski.
\newblock {\em Phys. Rev. A} 72, 012103 (2005).

\bibitem{Caban2006}
P. Caban and J. Rembieli{\'n}ski.
\newblock {\em Phys. Rev. A} 74, 042103 (2006).

\bibitem{Landulfo2009}
A. G. S. Landulfo and G. E. A. Matsas.
\newblock {\em Phys. Rev. A} 79, 044103 (2009).

\bibitem{Debarba2012}
T. Debarba and R. O. Vianna.
\newblock {\em Int. J. Quantum Inform.} 10, 1230003 (2012).

\bibitem{Saldanha2012(2)}
P. L. Saldanha and V. Vedral.
\newblock {\em Phys. Rev. A} 85, 062101 (2012).

\bibitem{Awschalom2013}
{D.D. Awschalom, L. C. Basset, A. S. Dzurak, E. L. Hu, and J. R. Petta}
\newblock {\em Science} 339, 1174 (2013).

\bibitem{Piazza2012}
A.~Di~Piazza, C.~Muller, K.~Z. Hatsagortsyan, and C.~H. Keitel.
\newblock {\em Rev. Mod. Phys.} 84, 1177 (2012).

\bibitem{Ahrens2012}
S.~Ahrens, H.~Bauke, C.~H. Keitel, and C.~Muller.
\newblock {\em Phys. Rev. Lett.} 109, 043601 (2012).

\bibitem{Malcolm}
M. Ludvigsen.
\newblock {\em General Relativity - A Geometric Approach}.
\newblock Cambridge University Press, 1999.

\bibitem{Wigner}
E.~Wigner.
\newblock {\em Ann. Math.} 40, 149 (1939).

\bibitem{Enk1994}
S.~J. van Enk and G.~Nienhuis.
\newblock {\em Europhys. Lett.} 25, 497 (1994).

\bibitem{Thomas1926}
L.~H. Thomas.
\newblock {\em Nature} 117, 514 (1926).

\end{thebibliography}
\end{document}